\documentclass[sigconf]{acmart}

\copyrightyear{2026}
\acmYear{2026}
\setcopyright{cc}
\setcctype{by-nc-nd}
\acmConference[SIGIR '26]{Proceedings of the 49th International ACM SIGIR Conference on Research and Development in Information Retrieval}{July 20--24, 2026}{Melbourne, VIC, Australia}
\acmBooktitle{Proceedings of the 49th International ACM SIGIR Conference on Research and Development in Information Retrieval (SIGIR '26), July 20--24, 2026, Melbourne, VIC, Australia}
\acmDOI{10.1145/3805712.3808475}
\acmISBN{979-8-4007-2599-9/2026/07}

\usepackage{multirow}

\usepackage{fancyhdr}
\fancypagestyle{firstpage}{
  \fancyhf{}
  \fancyhead[L]{\textit{Accepted at ACM SIGIR 2026, Melbourne, Australia.}}
  
}

\begin{document}

\title{LLM-Enhanced Topical Trend Detection at Snapchat}

\author{Hangqi Zhao}
\affiliation{%
  \institution{Snap Inc.}
  \city{Bellevue}
  \state{Washington}
  \country{USA}}
\email{hzhao@snapchat.com}

\author{Jay Li}
\authornote{Work performed while at Snap Inc.}
\affiliation{%
  \institution{Snap Inc.}
  \city{Palo Alto}
  \state{California}
  \country{USA}}
\email{wenjie.li@snapchat.com}

\author{Abhiruchi Bhattacharya}
\affiliation{%
  \institution{Snap Inc.}
  \city{Palo Alto}
  \state{California}
  \country{USA}}
\email{abhattacharya2@snapchat.com}

\author{Cong Ni}
\affiliation{%
  \institution{Snap Inc.}
  \city{Palo Alto}
  \state{California}
  \country{USA}}
\email{sni@snapchat.com}

\author{Jason Yeung}
\affiliation{%
  \institution{Snap Inc.}
  \city{New York}
  \state{New York}
  \country{USA}}
\email{jyeung@snapchat.com}

\author{Jinchao Ye}
\affiliation{%
  \institution{Snap Inc.}
  \city{New York}
  \state{New York}
  \country{USA}}
\email{jye2@snapchat.com}

\author{Kai Yang}
\affiliation{%
  \institution{Snap Inc.}
  \city{Palo Alto}
  \state{California}
  \country{USA}}
\email{kyang2@snapchat.com}

\author{Akshat Malu}
\affiliation{%
  \institution{Snap Inc.}
  \city{Bellevue}
  \state{Washington}
  \country{USA}}
\email{amalu@snapchat.com}

\author{Manish Malik}
\authornotemark[1]
\affiliation{%
  \institution{Snap Inc.}
  \city{Palo Alto}
  \state{California}
  \country{USA}}
\email{mmalik@snapchat.com}

\renewcommand{\shortauthors}{Hangqi Zhao et al.}

\begin{abstract}
Automatic detection of topical trends at scale is both challenging and essential for maintaining a dynamic content ecosystem on social media platforms. In this work, we present a large-scale system for identifying emerging topical trends on Snapchat, one of the world's largest short-video social platforms. Our system integrates multimodal topic extraction, time-series burst detection, and LLM-based consolidation and enrichment to enable accurate and timely trend discovery. To the best of our knowledge, this is the first published end-to-end system for topical trend detection on short-video platforms at production scale. Continuous offline human evaluation over six months demonstrates high precision in identifying meaningful trends. The system has been deployed in production at global scale and applied to downstream surfaces including content ranking and search, driving measurable improvements in content freshness and user experience.
\end{abstract}

\ccsdesc[500]{Computing methodologies~Computer vision}
\ccsdesc[500]{Computing methodologies~Natural language processing}
\ccsdesc[300]{Information systems~Data mining}
\ccsdesc[500]{Information systems~Recommender systems}
\ccsdesc[500]{Information systems~Document topic models}

\keywords{Trend Detection, Topic Detection, Large Language Models, Vision-Language Models, Short-Video Platforms, Social Media, Multimodality, Content Understanding, Recommender Systems}

\maketitle
\thispagestyle{firstpage}

\section{Introduction}

Topical trend detection plays a critical role in short-video and multimedia social platforms~\cite{allan1998tdt,weng2011event}. On platforms such as Snapchat, TikTok, and Instagram, these trends drive large-scale content creation and consumption patterns, closely interacting with modern sequential and short-video recommendation systems~\cite{kang2018self,zhou2018deep,zhang2023tiktok}. In particular, Snapchat is a major short-video platform where content is consumed through surfaces such as Stories and Spotlight. Timely identification of emerging trends enables better content surfacing, personalization, and creator participation, making the ability to recognize \emph{what is trending, where, and why} a key differentiator for user experience.

However, detecting topical trends on short-video platforms poses unique challenges compared to traditional text-based social networks. First, the large volume of multimodal content requires scalable systems for efficient processing. Second, trends often emerge subtly through weak signals distributed across communities, making \emph{early detection} difficult~\cite{allan1998tdt,cataldi2010emerging,mathioudakis2010twittermonitor}. Finally, the rapid lifecycle of trends on social media means that models must adapt in near real-time while maintaining robustness against noise, spam, and ambiguous topics. Prior research on event and trend detection in text streams~\cite{allan1998tdt,cataldi2010emerging}, social networks~\cite{weng2011event}, and multimodal retrieval~\cite{singh2021multimodal} provides useful foundations but often falls short in handling the \emph{scale}, \emph{heterogeneity}, and \emph{temporal sensitivity} of short-video data.

In this work, we present Snapchat's large-scale topical trend detection system powered by Large Language Models (LLMs). We consider topical trends as time-sensitive emerging topics characterized by noticeable increases in content creation, reflecting evolving platform-wide interest. To the best of our knowledge, this is the first published end-to-end system for topical trend detection on short-video platforms at production scale. Our system integrates multimodal topic extraction, time-series burst detection, and LLM-based consolidation and enrichment to enable low-latency, near–real-time trend discovery. Its effectiveness is demonstrated through continuous offline human evaluation and online A/B experiments, and the system has been deployed at global scale, driving measurable improvements in content discovery and freshness.

Our main contributions are summarized as follows:
\begin{itemize}
    \item A scalable end-to-end architecture for trend detection in short-video social networks, combining multimodal topic understanding with time-series burst modeling.
    \item An efficient LLM-enhanced approach for trend extraction, consolidation, and enrichment that balances semantic accuracy with computational efficiency at platform scale.
    \item Comprehensive evaluation and deployment results demonstrating the system's precision and measurable improvements in content timeliness and discovery on Snapchat's global platform.
\end{itemize}

\section{Related Work}

\textbf{Topical Trend Detection.}
Emerging topic detection has been extensively studied in information retrieval and social media analytics. Early work focused on textual streams, using statistical burst detection and topic modeling techniques to identify rapidly evolving themes~\cite{allan1998tdt,cataldi2010emerging}. Dynamic topic models and continuous-time extensions further modeled topic evolution over time~\cite{blei2006dynamic,wang2007continuous}. In social media settings, prior studies incorporated temporal and network signals to detect trends on platforms such as Twitter~\cite{weng2011event,mathioudakis2010twittermonitor}. More recent approaches leverage neural representations and transformer-based embeddings to improve semantic clustering and ranking of trending topics~\cite{vaswani2017attention,reimers2019sentence}. However, most existing methods focus on text-based data, while trend detection for short-video platforms remains underexplored due to multimodal complexity, sparse metadata, and noisy user-generated signals~\cite{li2019detecting}.\\
\textbf{Multimodal Understanding and LLMs.}
Vision-Language models (VLMs) such as CLIP and BLIP/BLIP2~\cite{clip,blip2} enable cross-modal alignment between images and text for visual tagging and representation learning. Large language models (LLMs) including GPT-4, Gemini, LLaMA, and Qwen~\cite{openai2023gpt4,team2023gemini,touvron2023llama2,bai2025qwen2_5_vl,qwen3vl}, along with instruction-tuned variants~\cite{ouyang2022instructgpt}, have advanced semantic abstraction and reasoning over unstructured data. Recent surveys further characterize their evolving capabilities~\cite{minaee2024llm_survey,matarazzo2025llm_insights}. Multimodal LLMs extend these advances by integrating visual and textual inputs for cross-modal reasoning in video understanding tasks~\cite{sun2019videobert,alayrac2022flamingo,liu2023videollama}. These developments underpin LLM-enhanced topic extraction and consolidation in short-video trend detection systems.

\section{Methodology}

\subsection{System Overview}

\begin{figure}[!htbp]
  \centering
  \includegraphics[width=\linewidth]{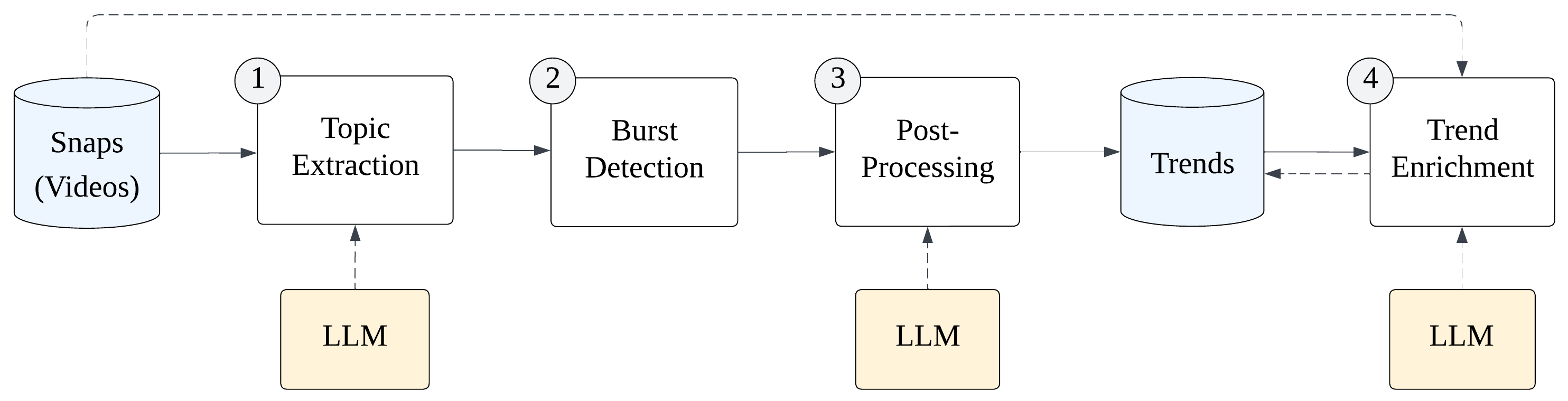}
  \caption{Overview of the trend detection system.}
  \Description{System architecture diagram showing data ingestion, keyword extraction, burst detection, topic consolidation, and integration layers.}
  \label{fig:snaptrend_pipeline}
\end{figure}

Figure~\ref{fig:snaptrend_pipeline} shows the end-to-end architecture of our system for early detection of emerging topical trends on Snapchat. The pipeline consists of four stages (Steps~1–4 in the figure) and runs on a regular cadence to keep detected trends fresh. In Step~1 (Topic Extraction), multimodal signals from Snaps—including visual tags, ASR transcripts, OCR text, and user-provided metadata—are processed by VLMs and LLMs to extract candidate topics. In Step~2 (Burst Detection), topic-level posting time series are analyzed to identify burst patterns indicative of emerging trends. These candidates are then refined in Step~3 (Post-processing) through LLM-assisted filtering and consolidation, which merges semantically similar topics and removes noise to produce high-confidence trends. Finally, in Step~4 (Trend Enrichment), trends are augmented with structured metadata such as summaries, descriptions, and categories, enabling integration with downstream recommendation, search, and discovery systems.

\subsection{Topic Extraction}
\begin{figure}[!htbp]
  \centering
  \includegraphics[width=\linewidth]{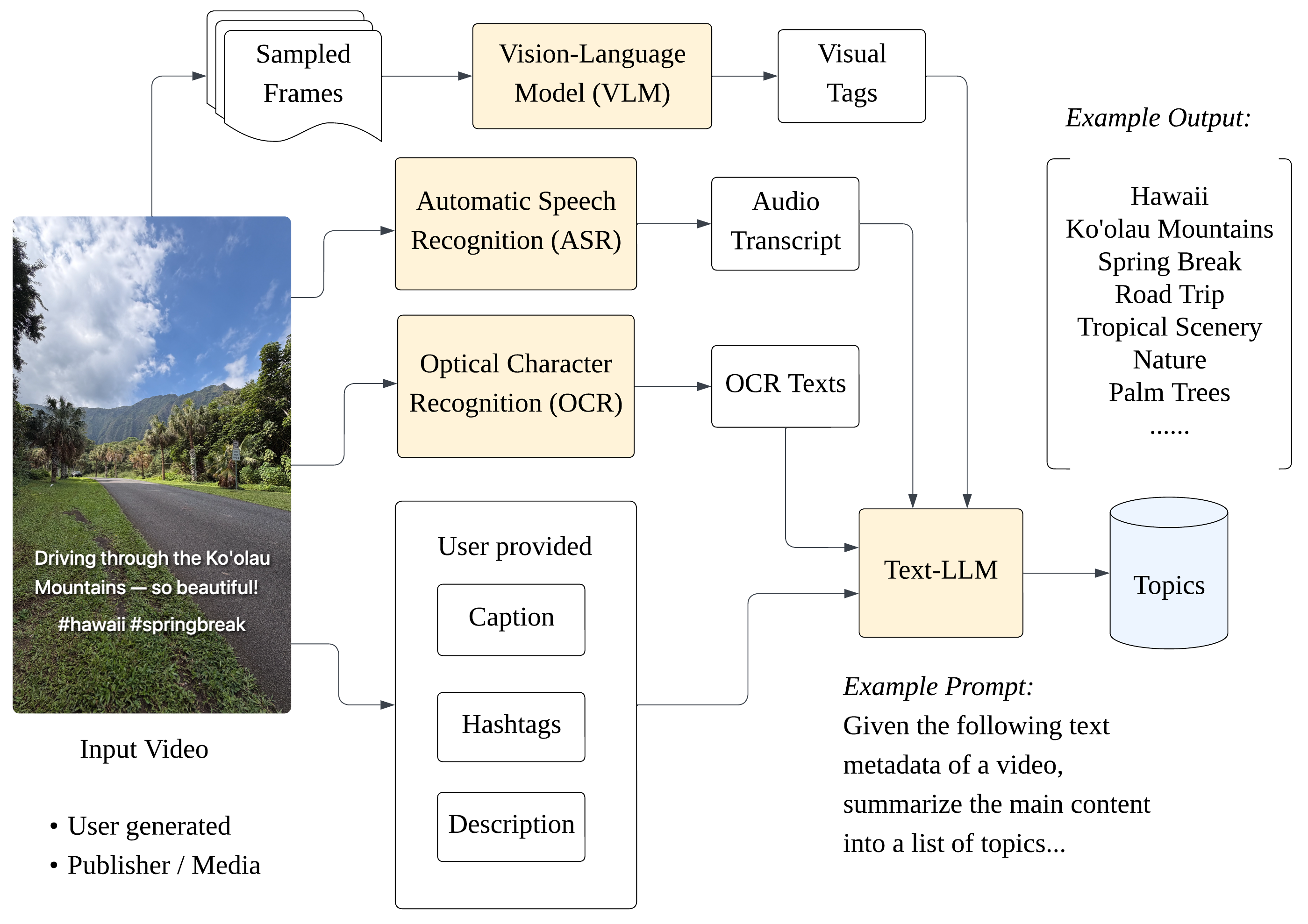}
  \caption{\textbf{Topic Extraction} module in the system.}
  \Description{A video frame is processed through a vision-language model to extract visual tags, captions, hashtags, and transcripts. The aggregated metadata is then summarized by an LLM to produce topic labels.}
  \label{fig:topic_extraction}
\end{figure}

The first stage of the pipeline extracts semantic topics from Snapchat short videos, including content from individual users and media publishers, where publisher videos often serve as high-quality sources for news, major events, and public figures. As shown in Figure~\ref{fig:topic_extraction}, each video is processed by a set of lightweight multimodal models to generate textual signals, including a Vision-Language Model (VLM) applied to sampled image frames to produce visual content tags from a fixed taxonomy based purely on visual signals, automatic speech recognition (ASR) for audio transcripts, and optical character recognition (OCR) for on-screen text. Since this stage operates at full platform scale, the models are chosen to balance accuracy with cost and latency constraints; For example, our system instantiates the VLM using BLIP2~\cite{blip2}, although other lightweight VLMs such as CLIP~\cite{clip} are also suitable for this stage.

The generated multimodal signals are combined with user-provided captions, hashtags, and descriptions and unified into a single textual representation. This representation is then passed to a text-only Large Language Model (LLM), which is prompted to summarize the video's main content into a concise set of free-form topical phrases or entities, rather than selecting from a predefined taxonomy. These extracted topics form the basis for downstream burst detection and trend aggregation.

\subsection{Burst Detection}
We model emerging trends as rapid increases in content creation and focus on detecting burst patterns in the number of \emph{unique users (UUs)} posting about a topic. Viewer-side signals are intentionally excluded to avoid bias introduced by downstream ranking mechanisms. Burst detection is applied to globally aggregated posting data by default, and the same formulation can be directly extended to country- or region-specific subsets for localized trend discovery.

To identify sudden surges in activity, we apply a multi-scale burst detection algorithm over topic-level posting time series. As a pre-filtering step, topics with fewer than $M$ UUs are removed to reduce long-tail noise and improve detection robustness.

\paragraph{Step 1: Moving Maxima and Averages.}
For each topic and time $t$, let $\text{num}_{\text{user}}(t)$ denote the number of UUs posting about the topic. In practice, $\text{num}_{\text{user}}(t)$ is aggregated over the most recent $T$ hours ending at $t$ rather than a single hour, to smooth short-term fluctuations.

To establish a robust baseline for each topic, we first compute the moving maximum over a sliding window of $N$ hours. This captures the peak posting activity within each window, providing a conservative reference that reflects recent high points rather than average behavior:
\[
\text{num}_{\text{max}}(N, t) = \max_{i = t - N + 1}^{t} \text{num}_{\text{user}}(i)
\]
We then compute the moving average of these maxima to obtain a smoothed baseline:
\[
\text{num}_{\text{max\_avg}}(N, t) = \frac{1}{N} \sum_{i = t - N + 1}^{t} \text{num}_{\text{max}}(N, i)
\]

\paragraph{Step 2: Lift Calculation.}
We define a lift score to measure deviation from recent baselines:
\[
\text{lift}(N, t) =
\frac{\text{num}_{\text{user}}(t)}
{\text{num}_{\text{max\_avg}}(N, t-1)}
\]
A lift $>1$ indicates activity exceeding recent historical levels.

\paragraph{Step 3: Trend Scoring.}
Lift values across multiple temporal windows are aggregated using a weighted harmonic mean:
\[
\text{trend\_score}(t) =
\frac{\sum_{N} w_{N}}
{\sum_{N} \frac{w_{N}}{\text{lift}(N, t)}},
\quad
w_{N} = e^{-\lambda N}
\]
This formulation emphasizes recent activity while penalizing weak signals at any timescale, ensuring that only topics exhibiting sustained multi-scale growth receive high trend scores. A higher trend score therefore indicates a stronger and more consistent trending signal.

\subsection{Post-processing}

After trend scores are computed for all topics, the Post-processing module refines and validates the results to ensure quality, compliance, and interpretability before downstream consumption. This stage applies a combination of rule-based and LLM-assisted operations as follows:\\
\textbf{Sensitive Filtering.} To comply with company policies and community guidelines, LLM prompts are augmented with policy-specific constraints to detect and remove topics that may contain sensitive, unsafe, or restricted content.\\
\textbf{Generality Filtering.} LLM-based filtering is applied to remove overly broad or generic topics (e.g., ``funny videos'', ``daily life'') that lack specificity, as they do not represent meaningful emerging trends despite potentially high posting volume.\\
\textbf{Precision Control.} To maintain high topic precision, results are filtered using adaptive thresholds on key metrics such as the trend score and user posting levels. Thresholds are dynamically tuned across different content types to balance coverage and accuracy.\\
\textbf{Topic Consolidation.} Semantically similar topics are merged through LLM-based clustering and canonicalization into single trend to eliminate duplicates and unify surface variants. Specifically, we instruct LLM to select the most representative topic as the final trend, among all similar and related topics. For example, topics such as ``World Cup 2026'', ``World Cup'', ``World Cup 2026 qualifiers'' may be merged by LLM into a single trend of ``World Cup 2026''. This ensures that the final trend set is compact, interpretable, and free from redundancy.

\subsection{Trend Enrichment}
\begin{figure}[!htbp]
  \centering
  \includegraphics[width=\linewidth]{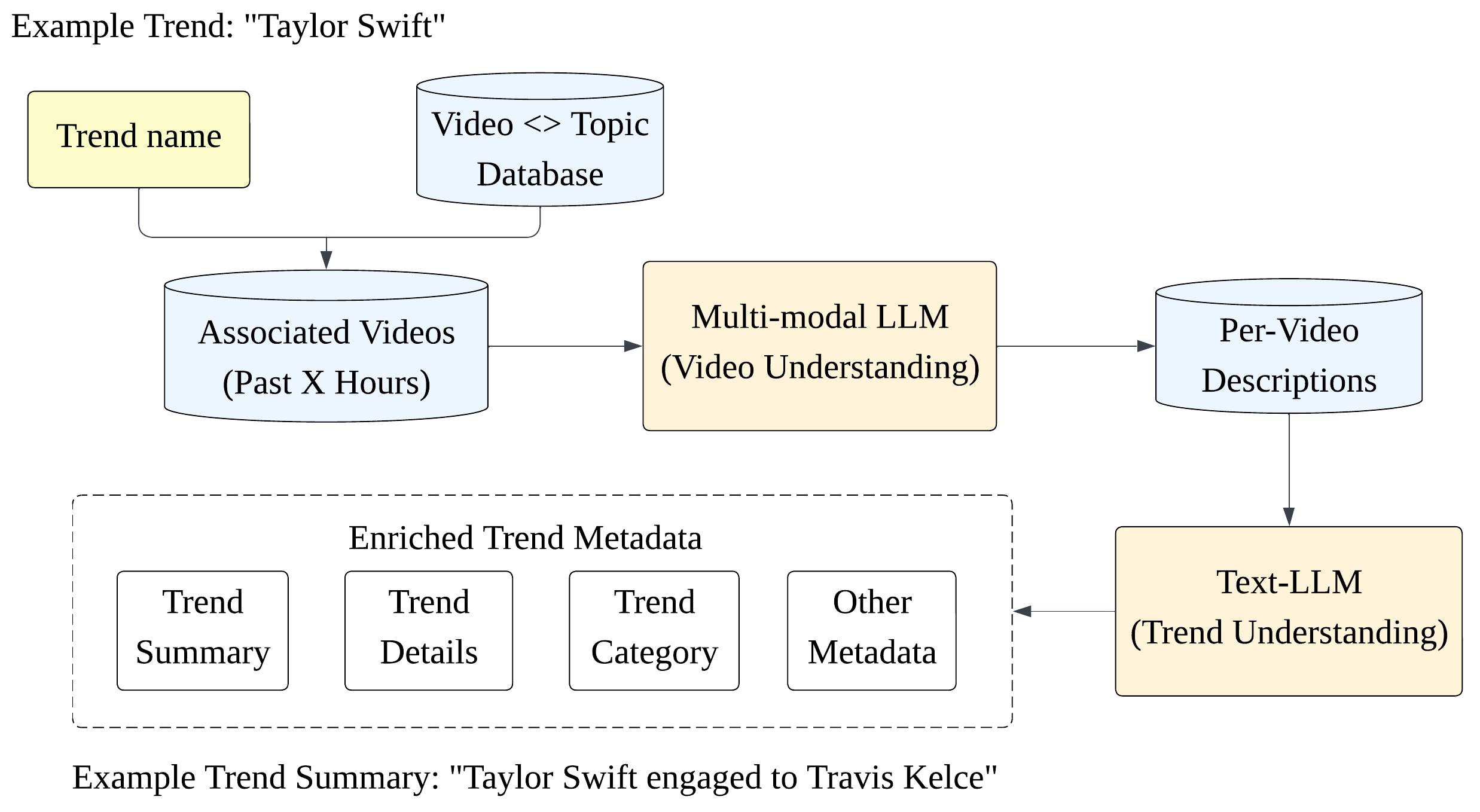}
  \caption{\textbf{Trend Enrichment} module in the system.}
  \Description{A flow diagram showing a trend connected to multiple videos, each processed by a multimodal LLM to produce video descriptions, which are then aggregated and summarized by a text-only LLM.}
  \label{fig:trend_enrichment}
\end{figure}

The Trend Enrichment module augments each detected trend with additional semantic context to improve interpretability and downstream usability. As shown in Figure~\ref{fig:trend_enrichment}, for each trend we collect a representative set of associated videos identified through the topic extraction module. Each video is processed by a Multimodal Large Language Model (MLLM), which analyzes visual, audio, and textual signals to produce a concise description of its main content. In our system, Gemini 2.0 Flash~\cite{team2023gemini} is used as the MLLM, although other proprietary or open-source models such as Qwen~\cite{qwen3vl} are also applicable. Unlike topic extraction, which operates at full platform scale, trend enrichment processes a smaller set of representative videos. This allows the use of more computationally intensive MLLMs to prioritize semantic accuracy over inference cost.

The generated video descriptions are then aggregated and passed to a text-only LLM for trend-level synthesis. The LLM produces a human-readable summary, extracts key details, assigns a canonical category (e.g., sports, entertainment, news), and generates structured metadata. An example of the enriched trend output is shown in Table~\ref{tab:trend_fields}.

\begin{table}[!htbp]
\centering
\small
\caption{Example fields in the enriched trend representation.}
\label{tab:trend_fields}
\begin{tabular}{ll}
\toprule
\textbf{Field} & \textbf{Description} \\
\midrule
Trend Name & A short word/phrase describing the trend \\
Detection Time & Timestamp when the trend is detected \\
Trend Score & Burst-based score indicating trend strength \\
Trend Summary & Concise high-level summary of the trend \\
Trend Details & Detailed explanation of the trend \\
Top Countries & Regions with the most trend postings \\
Trend Category & Content categories of the trend \\
\bottomrule
\end{tabular}
\end{table}

\section{Results}
\subsection{Offline Evaluation}

We conducted continuous offline human evaluation to assess the precision of trends detected by our system. From July to December 2025, random samples were reviewed weekly by an independent team of annotators, who labeled each trend (together with its detection time) as \emph{Correct} or \emph{Incorrect} based on whether it represented a meaningful emerging topic at that time.

Across 1{,}278 manually reviewed trends, our system achieves an overall precision of 92.8\%. The system also successfully captured major events, news stories, and widely discussed topics during the evaluation period, demonstrating its ability to identify meaningful real-world trends. We do not report recall due to the absence of a comprehensive global ground-truth dataset.

\subsubsection{Trend Score Sensitivity Analysis}

\begin{figure}[!htbp]
\centering
\includegraphics[width=0.9\linewidth]{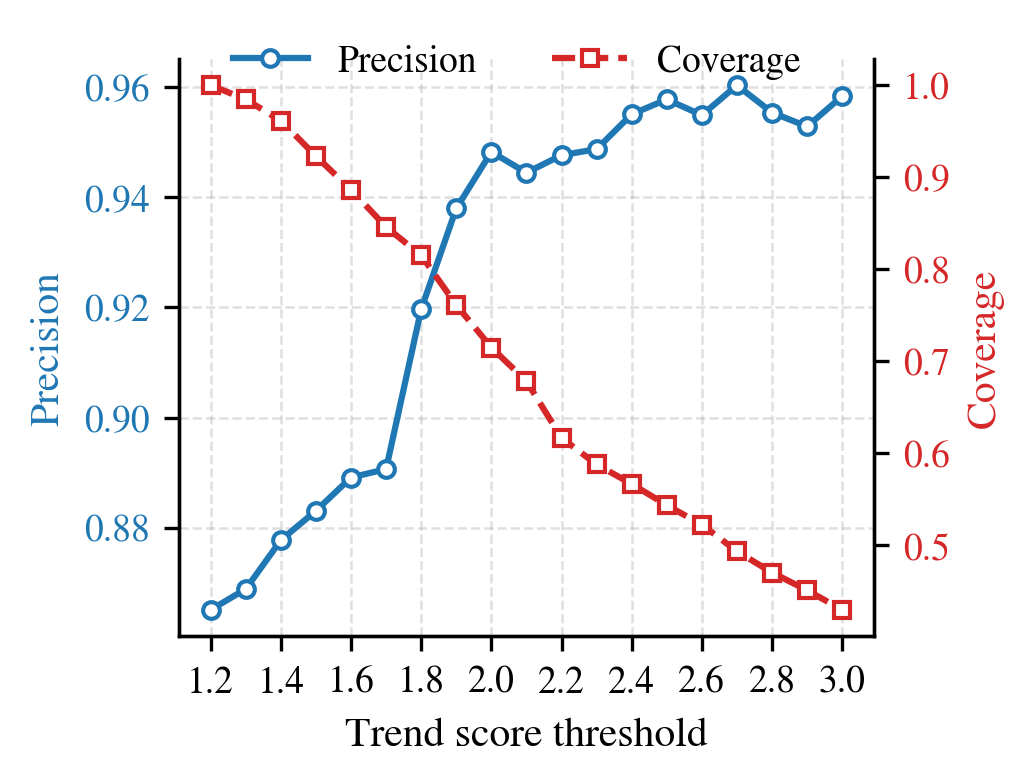}
\caption{Precision and coverage versus trend score threshold.}
\Description{A line chart showing precision increasing and coverage decreasing as the trend score threshold increases from 1.0 to 3.0.}
\label{fig:pc_curve}
\end{figure}

We analyze how the trend score threshold affects precision and coverage, as it determines which burst candidates are promoted to finalized trends. As shown in Figure~\ref{fig:pc_curve}, increasing the threshold improves precision while reducing coverage, reflecting the expected precision–coverage trade-off. This also demonstrates the effectiveness of the trend score. In production, we set the threshold to 1.8, achieving approximately 92\% precision. 

\subsection{Online A/B Testings}
We evaluated the real-world impact of our system through online A/B experiments on the Snapchat platform. Detected trends are integrated into multiple user-facing surfaces, most notably content ranking and content search.

\begin{figure}[!htbp]
\centering
\includegraphics[width=0.9\linewidth]{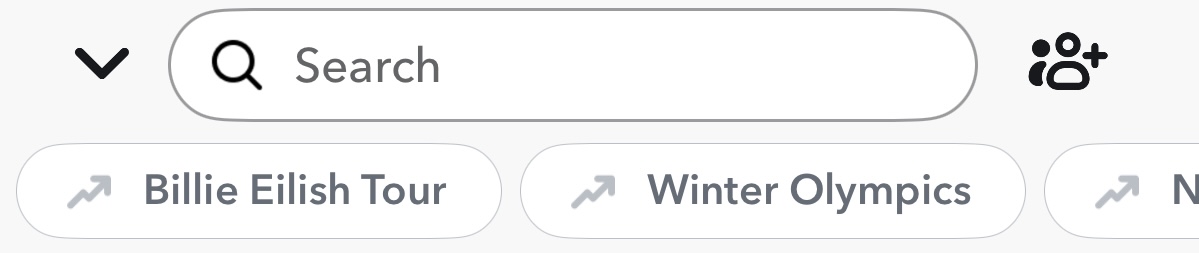}
\caption{Example search pretype UI}
\Description{A mobile screenshot showing trending topic suggestions in the Snapchat search bar before the user types a query.}
\label{fig:pretype}
\end{figure}

In content ranking, trends are used as a strategic signal to boost stories associated with detected trends, enabling timely content to surface more prominently while preserving overall recommendation quality. In content search, detected trends are surfaced as query suggestions in the search pretype experience (example shown in Figure~\ref{fig:pretype}), facilitating faster query formulation and improved discovery of emerging topics.

We conducted online A/B tests comparing baseline systems with trend-augmented variants for both ranking and search. Table~\ref{tab:ab_overall} summarizes the relative improvements across key categories. Trend-aware signals consistently improve \emph{content timeliness} by accelerating the exposure of newly emerging stories, leading to increased story views. In addition, trending content drives higher \emph{like rate} than non-trending content, indicating stronger user affinity with timely and culturally relevant topics. It also improves search abandonment rate and popular account open.

\begin{table}[!htbp]
\centering
\caption{Relative improvements from online A/B testing with trend detection integration.}
\label{tab:ab_overall}
\begin{tabular}{lp{4.1cm}c}
\toprule
\textbf{Surface} & \textbf{Metric} & \textbf{Improvement (\%)} \\
\midrule
\multirow{3}{*}{Ranking}
& Spotlight Story View (US)               & +0.86\% \\
& Non Friend Story View Time               & +0.15\% \\
& Content Like Rate                    & +11.5\% \\
\midrule
Ranking
& Content Timeliness (<1 day) & +1.89\% \\
\midrule
\multirow{3}{*}{Search}
& Impression Timeliness (<3 days) & +2.38\% \\
& PostType Abandonment Rate       & --1.28\% \\
& Popular Account Open (US)       & +26\% \\
\bottomrule
\end{tabular}
\end{table}

Based on consistent positive results from online experiments, our system has been fully launched and integrated into Snapchat's production systems serving its global user base, and continuously provides high-quality trend signals to ranking, search, and other content discovery surfaces.

\section{Conclusion}

We presented a production system for emerging topical trend detection on Snapchat, a large-scale short-video social media platform. Our system integrates multimodal topic extraction, time-series burst detection, and LLM-based post-processing and enrichment to enable timely and accurate trend discovery in dynamic content environments. This work demonstrates how large language models can be effectively combined with scalable information retrieval and temporal modeling pipelines in real-world systems. We hope the design choices and lessons learned from this work provide practical insights for building trend-aware retrieval and discovery systems in large-scale, multimodal platforms.

\section*{Acknowledgments}

We would like to thank Annie Zhang, Darby Haller, Vaishakh R, Sara Luo, Clark Ju, Jenn Velez, Lena Ingenillem-Woods, Linlin Chen, Mary Peng, Yuriana Zamora, Alex Farivar, and Chunhui Zhu for their valuable insights, feedback and support on this project.

\section*{Presenter Bio}
Hangqi Zhao is a Machine Learning Engineer in the Content organization at Snap Inc., where he works on content understanding, trend detection, large language models, and their applications in recommendation systems. Prior to joining Snap, he was a Senior Machine Learning Engineer at Twitter and an Applied Scientist at Amazon, focusing on large-scale ranking systems and natural language processing models. Hangqi holds a Ph.D. in Electrical and Computer Engineering from Rice University and has published widely.

\bibliographystyle{ACM-Reference-Format}
\balance
\bibliography{trend-detection-camera-ready}

@inbook{allan1998tdt,
author = {Allan, James},
title = {Introduction to topic detection and tracking},
year = {2002},
isbn = {0792376641},
publisher = {Kluwer Academic Publishers},
address = {USA},
booktitle = {Topic Detection and Tracking: Event-Based Information Organization},
pages = {1–16},
numpages = {16}
}

@inproceedings{cataldi2010emerging,
author = {Cataldi, Mario and Di Caro, Luigi and Schifanella, Claudio},
title = {Emerging topic detection on Twitter based on temporal and social terms evaluation},
year = {2010},
isbn = {9781450302203},
publisher = {Association for Computing Machinery},
address = {New York, NY, USA},
url = {https://doi.org/10.1145/1814245.1814249},
doi = {10.1145/1814245.1814249},
booktitle = {Proceedings of the Tenth International Workshop on Multimedia Data Mining},
articleno = {4},
numpages = {10},
location = {Washington, D.C.},
series = {MDMKDD '10}
}

@article{weng2011event, title={Event Detection in Twitter}, volume={5}, url={https://ojs.aaai.org/index.php/ICWSM/article/view/14102}, DOI={10.1609/icwsm.v5i1.14102}, number={1}, journal={Proceedings of the International AAAI Conference on Web and Social Media}, author={Weng, Jianshu and Lee, Bu-Sung}, year={2021}, month={Aug.}, pages={401-408} }

@article{singh2021multimodal,
  title={Multimodal event detection in social media videos},
  author={Singh, Karan and Kankanhalli, Mohan S.},
  journal={IEEE Transactions on Multimedia},
  year={2021}
}

@inproceedings{blei2006dynamic,
author = {Blei, David M. and Lafferty, John D.},
title = {Dynamic topic models},
year = {2006},
isbn = {1595933832},
publisher = {Association for Computing Machinery},
address = {New York, NY, USA},
url = {https://doi.org/10.1145/1143844.1143859},
doi = {10.1145/1143844.1143859},
booktitle = {Proceedings of the 23rd International Conference on Machine Learning},
pages = {113–120},
numpages = {8},
location = {Pittsburgh, Pennsylvania, USA},
series = {ICML '06}
}

@inproceedings{wang2007continuous,
author = {Wang, Chong and Blei, David and Heckerman, David},
title = {Continuous time dynamic topic models},
year = {2008},
isbn = {0974903949},
publisher = {AUAI Press},
address = {Arlington, Virginia, USA},
booktitle = {Proceedings of the Twenty-Fourth Conference on Uncertainty in Artificial Intelligence},
pages = {579–586},
numpages = {8},
location = {Helsinki, Finland},
series = {UAI'08}
}

@inproceedings{mathioudakis2010twittermonitor,
author = {Mathioudakis, Michael and Koudas, Nick},
title = {TwitterMonitor: trend detection over the twitter stream},
year = {2010},
isbn = {9781450300322},
publisher = {Association for Computing Machinery},
address = {New York, NY, USA},
url = {https://doi.org/10.1145/1807167.1807306},
doi = {10.1145/1807167.1807306},
booktitle = {Proceedings of the 2010 ACM SIGMOD International Conference on Management of Data},
pages = {1155–1158},
numpages = {4},
location = {Indianapolis, Indiana, USA},
series = {SIGMOD '10}
}

@inproceedings{vaswani2017attention,
 author = {Vaswani, Ashish and Shazeer, Noam and Parmar, Niki and Uszkoreit, Jakob and Jones, Llion and Gomez, Aidan N and Kaiser, \L ukasz and Polosukhin, Illia},
 booktitle = {Advances in Neural Information Processing Systems},
 editor = {I. Guyon and U. Von Luxburg and S. Bengio and H. Wallach and R. Fergus and S. Vishwanathan and R. Garnett},
 pages = {},
 publisher = {Curran Associates, Inc.},
 title = {Attention is All you Need},
 url = {https://proceedings.neurips.cc/paper_files/paper/2017/file/3f5ee243547dee91fbd053c1c4a845aa-Paper.pdf},
 volume = {30},
 year = {2017}
}

@inproceedings{reimers2019sentence,
    title = "Sentence-{BERT}: Sentence Embeddings using {S}iamese {BERT}-Networks",
    author = "Reimers, Nils  and
      Gurevych, Iryna",
    editor = "Inui, Kentaro  and
      Jiang, Jing  and
      Ng, Vincent  and
      Wan, Xiaojun",
    booktitle = "Proceedings of the 2019 Conference on Empirical Methods in Natural Language Processing and the 9th International Joint Conference on Natural Language Processing (EMNLP-IJCNLP)",
    month = nov,
    year = "2019",
    address = "Hong Kong, China",
    publisher = "Association for Computational Linguistics",
    url = "https://aclanthology.org/D19-1410/",
    doi = "10.18653/v1/D19-1410",
    pages = "3982--3992",
}

@INPROCEEDINGS{li2019detecting,
  author={Li, Quanzhi and Chao, Yang and Li, Dong and Lu, Yao and Zhang, Chi},
  booktitle={2022 IEEE International Conference on Big Data (Big Data)}, 
  title={Event Detection from Social Media Stream: Methods, Datasets and Opportunities}, 
  year={2022},
  volume={},
  number={},
  pages={3509-3516},
  doi={10.1109/BigData55660.2022.10020411}}

@INPROCEEDINGS {sun2019videobert,
author = { Sun, Chen and Myers, Austin and Vondrick, Carl and Murphy, Kevin and Schmid, Cordelia },
booktitle = { 2019 IEEE/CVF International Conference on Computer Vision (ICCV) },
title = {{ VideoBERT: A Joint Model for Video and Language Representation Learning }},
year = {2019},
volume = {},
ISSN = {},
pages = {7463-7472},
doi = {10.1109/ICCV.2019.00756},
url = {https://doi.ieeecomputersociety.org/10.1109/ICCV.2019.00756},
publisher = {IEEE Computer Society},
address = {Los Alamitos, CA, USA},
month =Nov}

@inproceedings{alayrac2022flamingo,
author = {Alayrac, Jean-Baptiste and Donahue, Jeff and Luc, Pauline and Miech, Antoine and Barr, Iain and Hasson, Yana and Lenc, Karel and Mensch, Arthur and Millicah, Katie and Reynolds, Malcolm and Ring, Roman and Rutherford, Eliza and Cabi, Serkan and Han, Tengda and Gong, Zhitao and Samangooei, Sina and Monteiro, Marianne and Menick, Jacob and Borgeaud, Sebastian and Brock, Andrew and Nematzadeh, Aida and Sharifzadeh, Sahand and Binkowski, Mikolaj and Barreira, Ricardo and Vinyals, Oriol and Zisserman, Andrew and Simonyan, Karen},
title = {Flamingo: a visual language model for few-shot learning},
year = {2022},
isbn = {9781713871088},
publisher = {Curran Associates Inc.},
address = {Red Hook, NY, USA},
booktitle = {Proceedings of the 36th International Conference on Neural Information Processing Systems},
articleno = {1723},
numpages = {21},
location = {New Orleans, LA, USA},
series = {NIPS '22}
}

@inproceedings{zhou2018deep,
author = {Zhou, Guorui and Zhu, Xiaoqiang and Song, Chenru and Fan, Ying and Zhu, Han and Ma, Xiao and Yan, Yanghui and Jin, Junqi and Li, Han and Gai, Kun},
title = {Deep Interest Network for Click-Through Rate Prediction},
year = {2018},
isbn = {9781450355520},
publisher = {Association for Computing Machinery},
address = {New York, NY, USA},
url = {https://doi.org/10.1145/3219819.3219823},
doi = {10.1145/3219819.3219823},
booktitle = {Proceedings of the 24th ACM SIGKDD International Conference on Knowledge Discovery \& Data Mining},
pages = {1059–1068},
numpages = {10},
location = {London, United Kingdom},
series = {KDD '18}
}

@INPROCEEDINGS {kang2018self,
author = { Kang, Wang-Cheng and McAuley, Julian },
booktitle = { 2018 IEEE International Conference on Data Mining (ICDM) },
title = {{ Self-Attentive Sequential Recommendation }},
year = {2018},
volume = {},
ISSN = {},
pages = {197-206},
doi = {10.1109/ICDM.2018.00035},
url = {https://doi.ieeecomputersociety.org/10.1109/ICDM.2018.00035},
publisher = {IEEE Computer Society},
address = {Los Alamitos, CA, USA},
month =Nov}

@inproceedings{blip2,
author = {Li, Junnan and Li, Dongxu and Savarese, Silvio and Hoi, Steven},
title = {BLIP-2: bootstrapping language-image pre-training with frozen image encoders and large language models},
year = {2023},
publisher = {JMLR.org},
booktitle = {Proceedings of the 40th International Conference on Machine Learning},
articleno = {814},
numpages = {13},
location = {Honolulu, Hawaii, USA},
series = {ICML'23}
}

@misc{clip,
      title={Learning Transferable Visual Models From Natural Language Supervision}, 
      author={Alec Radford and Jong Wook Kim and Chris Hallacy and Aditya Ramesh and Gabriel Goh and Sandhini Agarwal and Girish Sastry and Amanda Askell and Pamela Mishkin and Jack Clark and Gretchen Krueger and Ilya Sutskever},
      year={2021},
      eprint={2103.00020},
      archivePrefix={arXiv},
      primaryClass={cs.CV},
      url={https://arxiv.org/abs/2103.00020}, 
}

@misc{qwen3vl,
      title={Qwen3-VL Technical Report}, 
      author={Shuai Bai and Yuxuan Cai and Ruizhe Chen and Keqin Chen and Xionghui Chen and Zesen Cheng and Lianghao Deng and Wei Ding and Chang Gao and Chunjiang Ge and Wenbin Ge and Zhifang Guo and Qidong Huang and Jie Huang and Fei Huang and Binyuan Hui and Shutong Jiang and Zhaohai Li and Mingsheng Li and Mei Li and Kaixin Li and Zicheng Lin and Junyang Lin and Xuejing Liu and Jiawei Liu and Chenglong Liu and Yang Liu and Dayiheng Liu and Shixuan Liu and Dunjie Lu and Ruilin Luo and Chenxu Lv and Rui Men and Lingchen Meng and Xuancheng Ren and Xingzhang Ren and Sibo Song and Yuchong Sun and Jun Tang and Jianhong Tu and Jianqiang Wan and Peng Wang and Pengfei Wang and Qiuyue Wang and Yuxuan Wang and Tianbao Xie and Yiheng Xu and Haiyang Xu and Jin Xu and Zhibo Yang and Mingkun Yang and Jianxin Yang and An Yang and Bowen Yu and Fei Zhang and Hang Zhang and Xi Zhang and Bo Zheng and Humen Zhong and Jingren Zhou and Fan Zhou and Jing Zhou and Yuanzhi Zhu and Ke Zhu},
      year={2025},
      eprint={2511.21631},
      archivePrefix={arXiv},
      primaryClass={cs.CV},
      url={https://arxiv.org/abs/2511.21631}, 
}

@article{openai2023gpt4,
  title={GPT-4 Technical Report},
  author={{OpenAI}},
  journal={arXiv preprint arXiv:2303.08774},
  year={2023}
}

@article{touvron2023llama2,
  title={LLaMA 2: Open Foundation and Fine-Tuned Chat Models},
  author={Touvron, Hugo and others},
  journal={arXiv preprint arXiv:2307.09288},
  year={2023}
}

@article{team2023gemini,
  title={Gemini: A Family of Highly Capable Multimodal Models},
  author={{Google DeepMind}},
  journal={arXiv preprint arXiv:2312.11805},
  year={2023}
}

@inproceedings{ouyang2022instructgpt,
  title={Training Language Models to Follow Instructions with Human Feedback},
  author={Ouyang, Long and others},
  booktitle={Advances in Neural Information Processing Systems (NeurIPS)},
  year={2022}
}

@article{liu2023videollama,
  title={Video-LLaMA: An Instruction-tuned Audio-Visual Language Model for Video Understanding},
  author={Liu, Hang and others},
  journal={arXiv preprint arXiv:2306.02858},
  year={2023}
}

@article{minaee2024llm_survey,
  title={Large Language Models: A Survey},
  author={Minaee, Shervin and Mikolov, Tomas and Nikzad, Narjes and Chenaghlu, Meysam and Socher, Richard and Amatriain, Xavier and Gao, Jianfeng},
  journal={arXiv preprint arXiv:2402.06196},
  year={2025}
}

@article{matarazzo2025llm_insights,
  title={A Survey on Large Language Models with Some Insights on their Capabilities and Limitations},
  author={Matarazzo, Andrea and Torlone, Riccardo},
  journal={arXiv preprint arXiv:2501.04040},
  year={2025}
}

@article{bai2025qwen2_5_vl,
  title={Qwen2.5-VL Technical Report},
  author={Bai, Shuai and Chen, Keqin and Liu, Xuejing and Wang, Jialin and others},
  journal={arXiv preprint arXiv:2502.13923},
  year={2025}
}

@article{zhang2023tiktok,
  title={Understanding Short-Video Recommendation at Scale: Modeling User Interest Evolution in TikTok},
  author={Zhang, X. and others},
  journal={arXiv preprint arXiv:2305.xxxxx},
  year={2023}
}

\end{document}